\documentclass{osa-article}
\graphicspath{{figs/}}

\journal{oe}

\articletype{Research Article}

\begin{document}

\title{Passive all-optical synchronization for polarization-maintaining ultrafast fiber lasers}

\author{Kun Huang,\authormark{1} Jing Zeng,\authormark{1} Jiwei Gan,\authormark{1} Qiang Hao,\authormark{1} Ming Yan,\authormark{2} and Heping Zeng\authormark{1,2,*}}

\address{\authormark{1}Shanghai Key Laboratory of Modern Optical System, and Engineering Research Center of Optical Instrument and System, Ministry of Education, School of Optical Electrical and Computer Engineering, University of Shanghai for Science and Technology, Shanghai 200093, China\\
\authormark{2}State Key Laboratory of Precision Spectroscopy, East China Normal University, Shanghai 200062, China}

\email{\authormark{*}hpzeng@phy.ecnu.edu.cn} 



\begin{abstract}
We have proposed and implemented for the first time to our best knowledge a passive and all-optical pulse synchronization for polarization-maintaining ultrafast fiber lasers. Specifically, the synchronization system was comprised of two independent Yb-doped and Er-doped mode-locked fiber lasers in a master-slave configuration. Master pulses were injected into the slave laser cavity consisting of a nonlinear amplifying loop mirror, which provided an effective fast intensity modulator due to the periodic introduction of nonreciprocal phase difference. As a result, robust and tight timing synchronization was achieved with a cavity mismatch tolerance of 800 $\mu$m and a relative timing jitter of 26 fs within 1-MHz bandwidth. In combination with all-polarization-maintaining structure of fiber lasers, long-term stable operation was demonstrated over 12 hours without the need of temperature stabilization and vibration isolation. The implemented synchronous laser system could find immediate applications such as pump-probe microscopy, two-color spectroscopy and nonlinear frequency mixing.
\end{abstract}

\section{Introduction}
Stable and precise synchronization of optical pulses plays an essential role in ultrafast science and technology. In particular, synchronized pulses with disparate colors are widely used in many applications, including time-resolved imaging and spectroscopy, nonlinear frequency conversion, pump-probe investigations, coherent pulse synthesis, and precise timing distribution \cite{Fischer2016, Kim2010, Shelton2001, Foreman2007}.  Up to date, timing synchronization between two independent ultrafast lasers has been realized in various schemes. In pioneering works, active synchronization was demonstrated by using techniques of phase locked loops \cite{Shelton2002} and balanced optical cross correlator \cite{Miura2002}, which ultimately led to an ultralow residual jitter with attosecond precision \cite{Schibli2003, Xin2017, Tian2017}. To mitigate the stringent requirement of complicated feed-back system and high-speed electronics, passive synchronization was proposed and implemented based on the nonlinear optical cross-phase modulation (XPM) \cite{Furst1996, Wei2002, Zhu2005} in a shared laser cavity, where the high-bandwidth feedback was provided by the instantaneous nonlinear response. 

Over the past years, the XPM-based passive synchronization has attracted increasing attention in fiber lasers due to enhanced nonlinear interaction within the confined and elongated transverse mode of the fiber waveguide \cite{Rusu2004A, Rusu2004B}. Compared with solid-state lasers, fiber lasers also offer the advantages of low cost, compactness, and higher efficiency.  Moreover, a so-called master-slave configuration can readily be deployed to obtain the sufficient nonlinearity for self-synchronization \cite{Yoshitomi2006, Zhou2009, Li2009, Huang2012, Yoshitomi2014}, which enables to overcome limitations in the shared intra-cavity scheme, for instance unstable gain competition, poor spectral tunability, and mutual cross-talk. Recently, active-passive hybrid approach for fiber lasers was investigated by combining slow cavity feedback and fast passive self-locking, which resulted in a long-term stability and a low timing error \cite{Tsai2013}. 

However, in the aforementioned master-slave architecture, the slave fiber lasers were typically constructed with non-polarization-maintaining fibers, which is required to implement synchronous mode-locking based on nonlinear polarization rotation \cite{Gu2010, Wu2014}. In this case, the synchronization system inevitably suffers from running instability due to ambient perturbation on polarization states, such as temperature fluctuation, mechanical vibration and pressure change. Additionally, the initiation of synchronous operation usually needs careful tuning of polarization states in the slave laser cavity and polarization optimization on injected master pulses. Consequently, self-starting and long-term stable synchronization may be hard to retained in practice. To alleviate the disturbing effects, the fiber laser should be installed into a foamed polystyrene box \cite{Yoshitomi2006} or be controlled with an active fashion \cite{Yoshitomi2014}, thus increasing the complexity of the whole system.

\begin{figure}[b!]
\centering
\includegraphics[width=0.70\columnwidth]{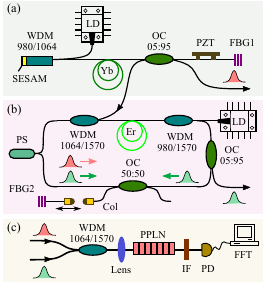}
\caption{Experimental setup for all-polarization-maintaining synchronization system. (a) Master Yb-doped mode-locked fiber laser. (b) Injection of master pulses into the slave Er-doped fiber laser cavity. The synchronous mode-locking is initiated by the additional non-reciprocal phase shift induced by the cross-phase modulation in the nonlinear amplifying loop mirror.  (c) Schematic for the measurement of relative timing jitter. LD: Laser diode; SESAM: semiconductor saturable absorber mirror; WDM: wavelength division multiplexer;  Yb: ytterbium-doped gain fiber;  OC: output coupler; PZT: piezoelectric transducer; FBG: fiber Bragg grating; PS: phase shifter; Er: erbium-doped gain fiber; Col: collimator; PPLN: periodically-poled lithium niobate crystal; IF: interference filter; PD: photodiode; FFT: fast Fourier transform.}
\label{fig1}
\end{figure}

Here, we proposed and implemented, for the first time to the best of our knowledge, an all-polarization-maintaining synchronization system for ultrafast fiber lasers in a master-slave layout.  Specifically, the optical master pulses from an Yb-doped mode-locked fiber laser were fed into a nonlinear amplifying loop mirror in the slave Er-doped fiber laser cavity, which periodically introduced non-reciprocal phase shift difference by the nonlinear cross-phase modulation. As a result, synchronous mode-locking of the slave laser could be obtained with a large tolerance of cavity mismatch up to 800 $\mu$m. Additionally, the relative timing jitter was measured by cross-correlation technique, revealing a root-mean-square value of 26 fs in a Fourier frequency range from 0.1 Hz to 1 MHz. The robust and tight synchronization could be launched in a plug-and-play way, and maintained more than 12 hours with a fraction frequency error of $5\times 10^{-10}$. The self-starting, highly precise and long-term stable synchronization system could promote its use in broader practical applications.

\section{Experimental setup}
\begin{figure}[b!]
\centering
\includegraphics[width=0.8\columnwidth]{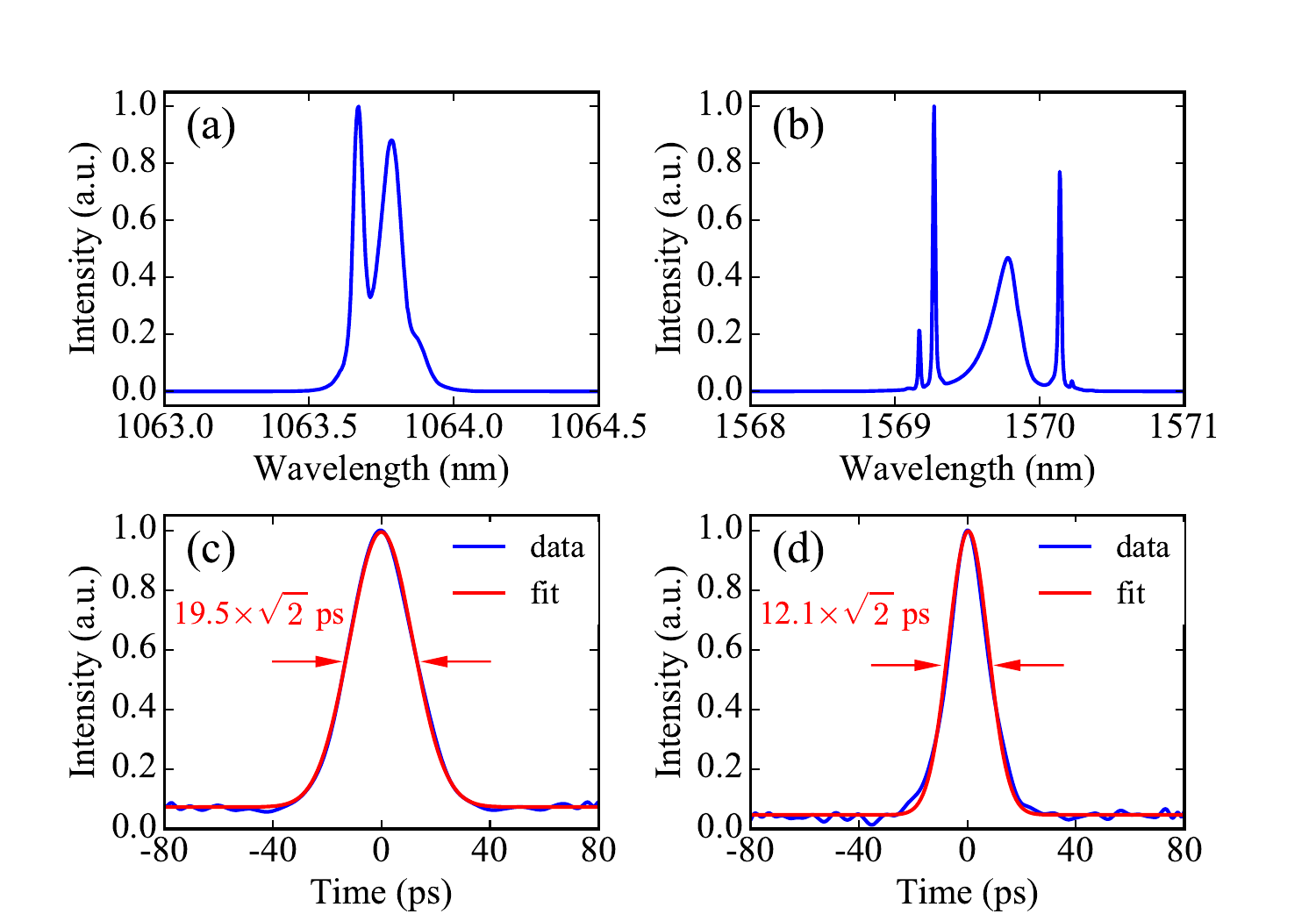}
\caption{Optical spectra of master (a) and slave (b) pulses. Autocorrelation traces for master (c) and slave (d) pulses.}
\label{fig2}
\end{figure}

Figure \ref{fig1} shows the experimental setup for the synchronized fiber laser system in a polarization-maintaining (PM) master-slave structure, which is comprised of an Yb-doped fiber laser (YDFL) as the master laser and an Er-doped fiber laser (EDFL) as the slave one. The master fiber laser is based on an all-PM linear cavity, where two ends are terminated with a semiconductor saturable absorber mirror (SESAM) and a fiber Bragg grating (FBG) with a bandwidth of 0.2 nm. Stable passive mode-locking can simply be obtained by increasing the pump power around 40 mW. The resulting pulses are then tapped bidirectionally through a $2 \times 2$ optical coupler with a splitting ration of 5:95. Figure \ref{fig2}(a) gives the measured optical spectrum centered at 1063.8 nm. The corresponding pulse duration of 19.5 ps can be deduced from the measured autocorrelation trace as shown in Fig. \ref{fig2}(c). Note that a piezoelectric transducer is glued with the intra-cavity fiber for actively stabilizing the cavity length, which will be elaborated later for the characterization of the synchronization system.

The slave fiber laser consists of a nonlinear amplifying loop mirror (NALM), which serves as a fast artificial saturable absorber. The related mechanism relies on the intensity-dependent nonlinear phase shift difference between clockwise and anticlockwise directions in the Sagnac loop \cite{Fermann1990}. It is thus imperative to accumulate sufficient phase difference to initiate mode-locked operation. For this purpose, the Er-doped gain fiber is placed asymmetrically relative to the $2 \times 2$ balanced optical coupler. Additionally, a $\pi$/2 non-reciprocal phase shifter is used to further reduce the mode-locking threshold \cite{Jiang2016, Chen2017, Hansel2017}. At one port of of the optical coupler, a 1-nm-bandwidth FBG is connected to effectively provide an end mirror and a spectral filter. Under a pump power above 280 mW, self-starting mode-locked pulses can be obtained with a pulse duration of 12.1 ps as shown in Fig. \ref{fig2}(d). The corresponding spectrum is centered at 1569.8 nm, and shows typical dispersive soliton-induced Kelly sidebands \cite{Chen2017}. The cavity length can be manually tuned by translating one of the two collimators in the free space as shown in Fig. \ref{fig1}(b).

\begin{figure}[b!]
\centering
\includegraphics[width=0.75\columnwidth]{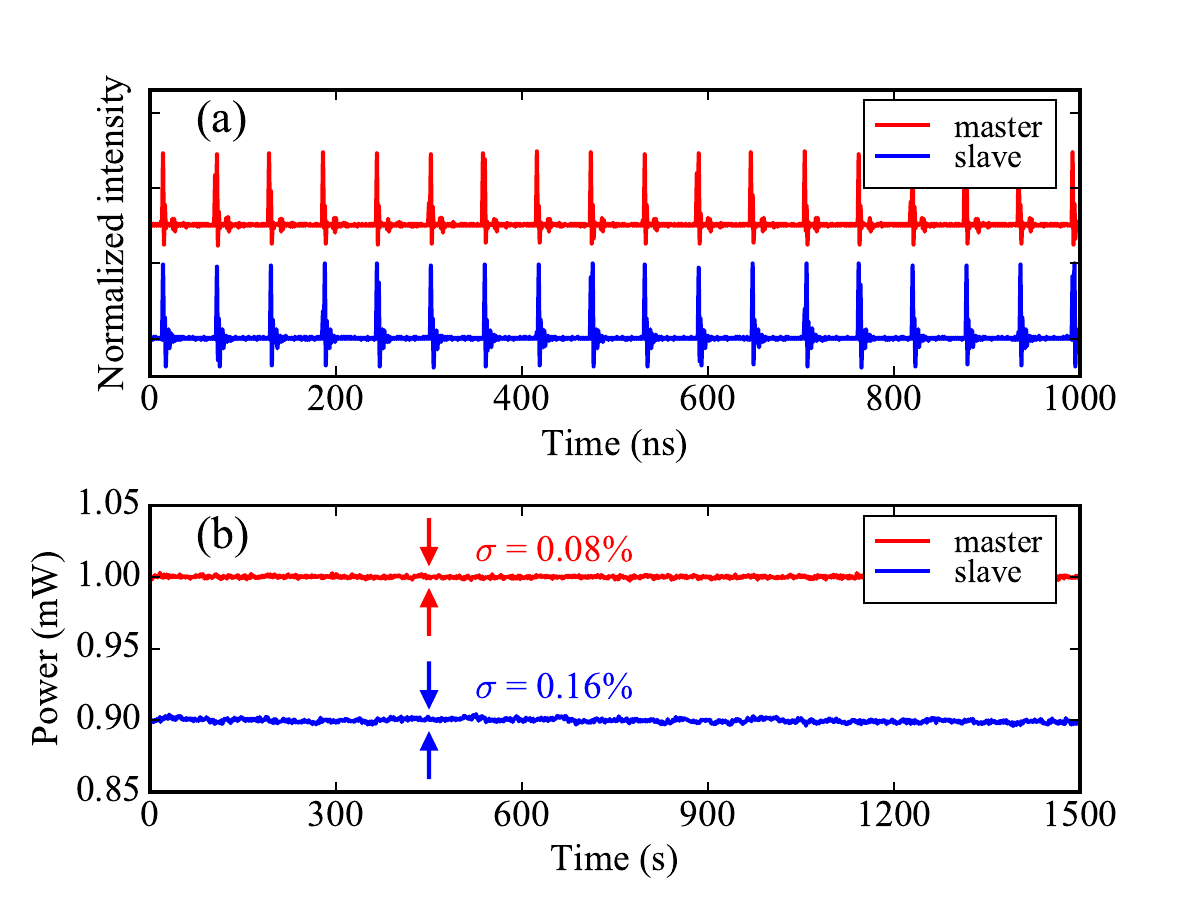}
\caption{Pulse trains (a) and power stability (b) for synchronized master and slave pulses. $\sigma$ indicates the relative fluctuation.}
\label{fig3}
\end{figure}

Now the two master-slave lasers are ready for synchronization. The master pulses at 1064 nm after an Yb-doped fiber amplifier (YDFA) are injected into the slave laser cavity via a wavelength division multiplexer (WDM). By carefully match the slave cavity length, synchronous mode-locked pulses can be observed as shown in Fig. \ref{fig3}(a). More specifically, the injected master pulses will induce additional phase shift in one direction of the NALM due to the cross-phase modulation, which provides a high-speed Kerr-based intensity modulation. Thanks to the all-PM design, the synchronization can be launched immediately with the light injection, without the need of any polarization state optimization. Figure \ref{fig3}(b) gives the measured relative power fluctuation after a linear polarizer, showing an excellent stability of 0.08\% and 0.16\% for master and slave laser outputs, respectively. 

\section{Results and discussion}
\begin{figure}[b!]
\centering
\includegraphics[width=0.65\columnwidth]{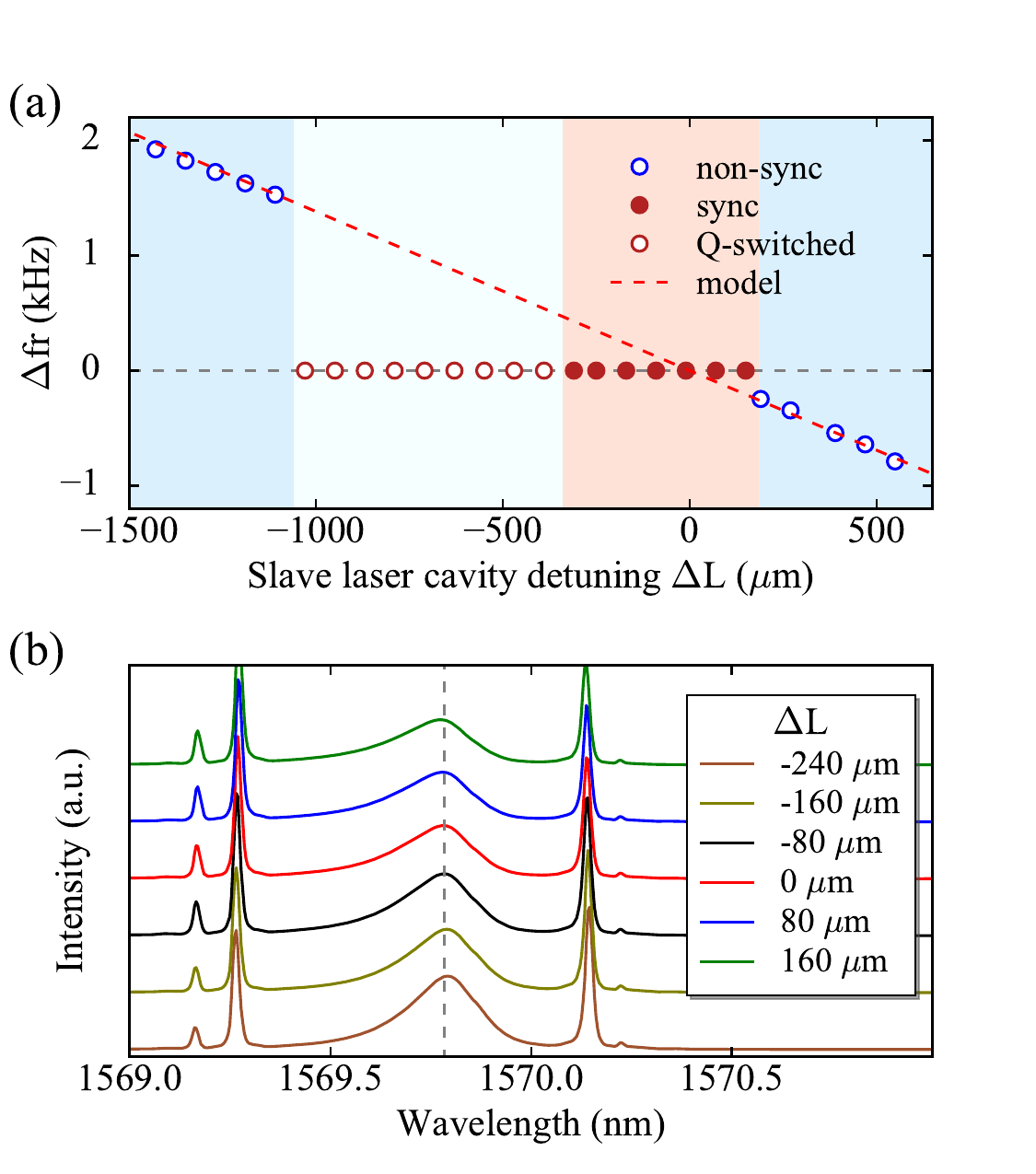}
\caption{(a)  Measured repetition rate of the slave laser as a function of its cavity-length variation. Red solid dots and blue open circles indicate synchronized and non-synchronized regimes, respectively. The intermediate Q-switched mode-locked regime is represented with red open circles. The dashed line is given by a theoretical mode in the absence of synchronization. (b) Optical spectra of the slave laser versus the cavity-length detuning.}
\label{fig4}
\end{figure}

Next we turn to characterize the implemented synchronization system. Figure \ref{fig4}(a) shows the cavity detuning diagram for repetition-rate changes when the slave laser operates at a pump power of 354 mW above the mode-locking threshold. The tolerance of the cavity-length mismatch is found to be about 460 $\mu$m, which corresponds to 635-Hz variation of free spectral range at a center repetition rate of 20.3 MHz. When the cavity-length change is away from the locking range, the repetition rate will jump to the value defined by a free-running laser, thus showing a typical linear dependence on the cavity detuning. Notably, an intermediate Q-switched mode-locked state is observed in the negative detuning regime, which indicates that the injected pulses impose a significant influence on the free-running slave laser. In addition, the negative capture range is slightly larger than the one in the positive regime, which may be ascribed to the intrinsic asymmetric structure of NALM in the slave laser cavity. Figure \ref{fig4}(b) shows the recorded spectra of synchronized slave pulses in various cavity-length detunings. The central wavelength is barely changed, which is in contrast to the previous master-slave technique based on nonlinear polarization rotation \cite{Rusu2004B, Yoshitomi2006, Tsai2013}. This feature of stable spectral positioning during the synchronization operation is favorable for applications requiring precise working wavelengths, such as coherent anti-Stokes Raman spectroscopy. 

\begin{figure}[t!]
\centering
\includegraphics[width=0.8\columnwidth]{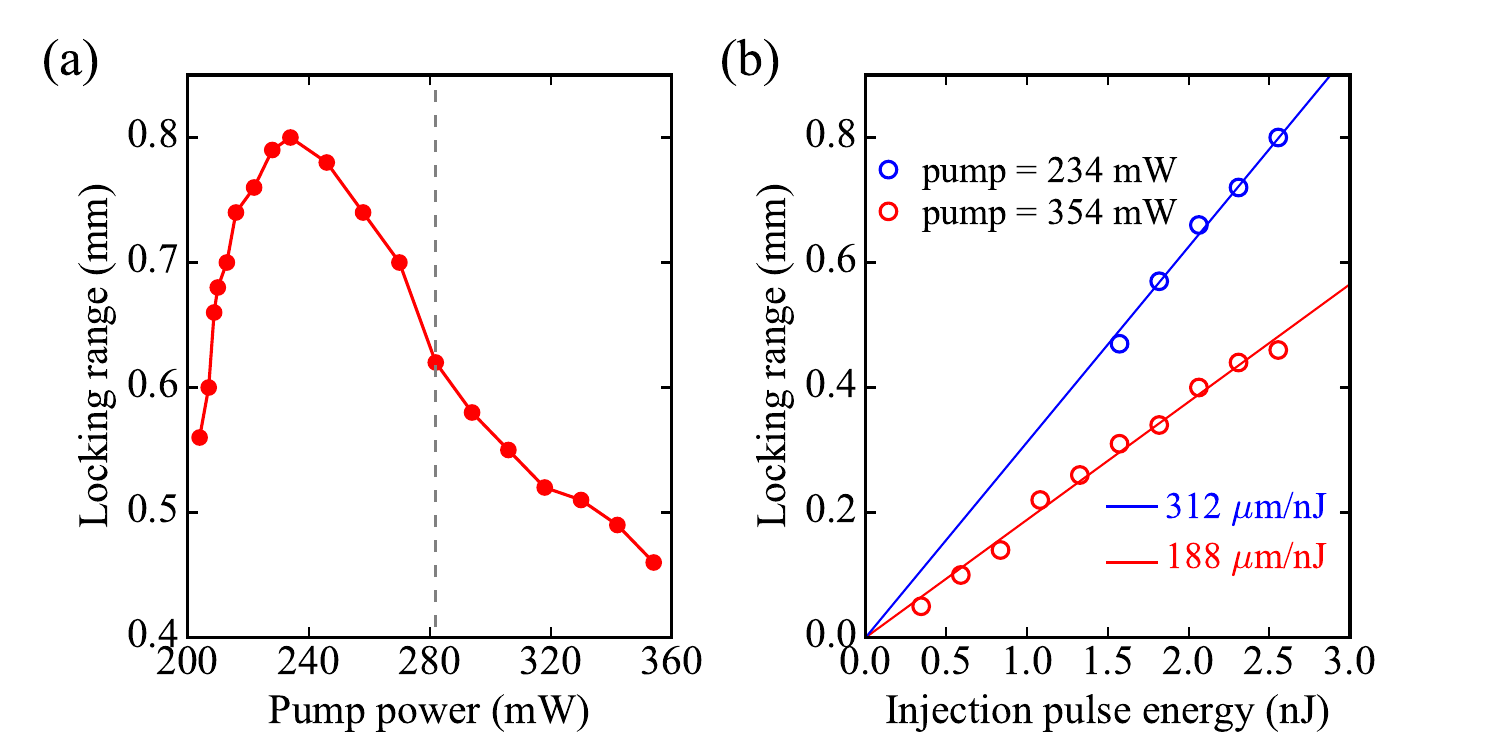}
\caption{(a) Cavity mismatch tolerance versus pump power of the slave laser. The dashed line indicates the mode-locking threshold in the free-running condition. (b) Cavity mismatch tolerance versus injection pulse energy from the master laser for above- and sub-threshold cases. The solid lines are given by linear fits. }
\label{fig5}
\end{figure}

In the following, we will further investigate two main factors that affect the locking range of pulse synchronization. In our experiment, we found that the locking range could be significantly increased by running the slave laser below the mode-locking threshold. In particular, the maximum tolerance of the cavity detuning reaches 800 $\mu$m at an optimum pump power of 234 mW as shown in Fig. \ref{fig5}(a). In this case, the required phase difference in NALM for mode locking is compensated by the Kerr-based phase shift by injected pulses. The other important factor to improve the capture range is the master-injecting pulse energy. As presented in Fig. \ref{fig5}(b), the range of the cavity-length tolerance increases linearly with augmenting the pulse energy. The fitted slopes are found to be 188 $\mu$m/nJ and 312 $\mu$m/nJ for two specific cases above and below the mode-locking threshold, respectively. Note that in the sub-threshold case, the synchronous mode-locking can only be obtained with sufficiently large injection pulse energy as expected. 

\begin{figure}[b!]
\centering
\includegraphics[width=0.65\columnwidth]{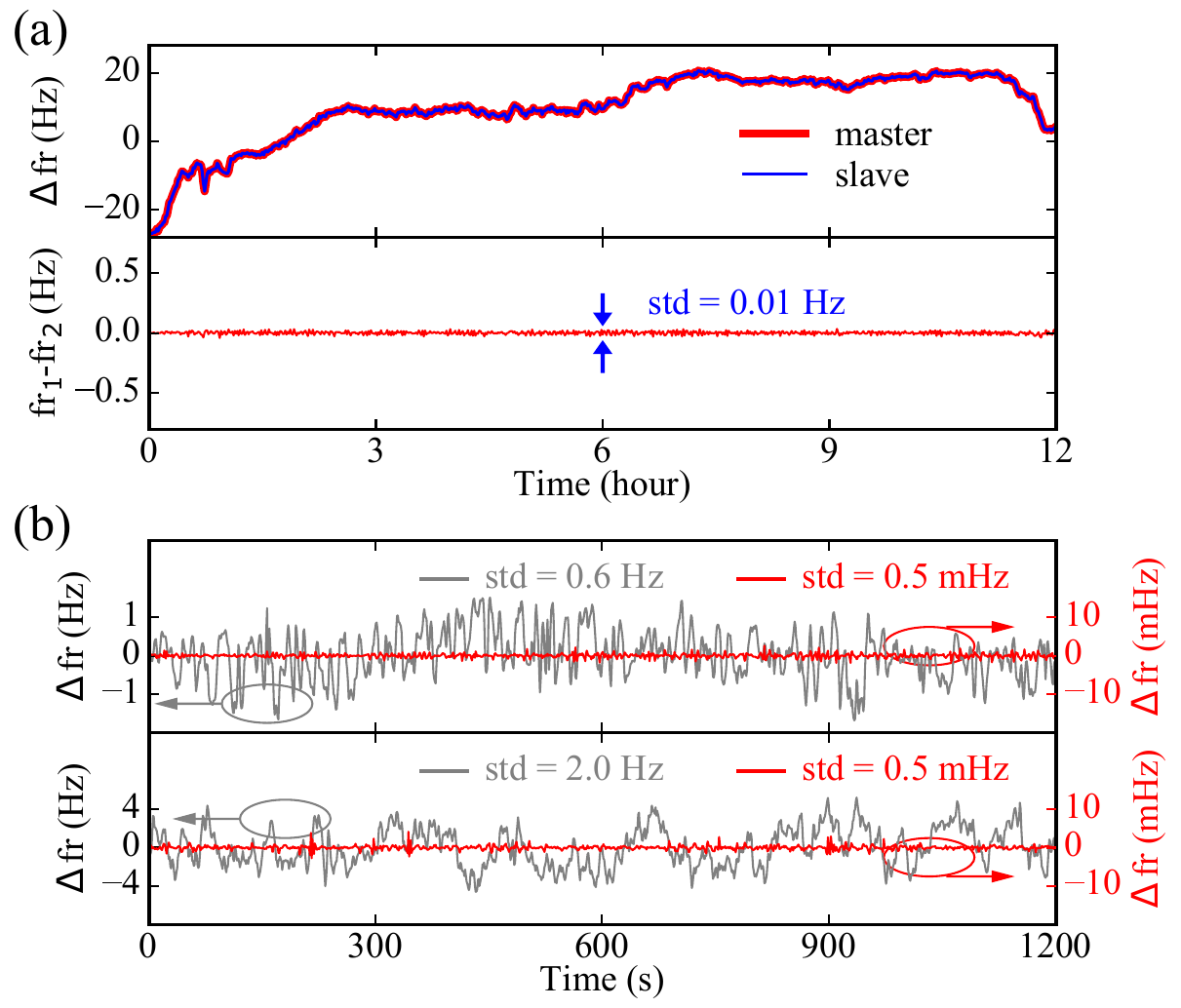}
\caption{(a) Repetition-rate changes for master and slave pulses within 12 hours (top panel) and their discrepancy (bottom panel). (b) Repetition-rate fluctuations of master laser with and without active cavity-length locking (top panel). Repetition-rate fluctuations of slave laser with and without synchronization (bottom panel).}
\label{fig6}
\end{figure}

In order to demonstrate the long-term stability, we have recorded the repetition rates of two synchronized fiber lasers by using digital frequency counters (Tektronix FCA3103). Figure \ref{fig6}(a) gives the result for an acquisition time as long as 12 hours, which shows a frequency discrepancy of 0.01 Hz in standard deviation. The corresponding frequency error relative to repetition rate is calculated to be $5 \times 10^{-10}$. On the other hand, the overall frequency difference for individual fiber laser is about 45 Hz, which is mainly due to the indoor-temperature fluctuation of 1.5 $^\circ$C. Considering a maximum capture range of 800 $\mu$m, our synchronization system could in principle resist a temperature variation of 37 $^\circ$C. Assisted with the PM mode-locking mechanism, the presented system could potentially operate outdoors beyond the laboratory environment. 

We then further investigate the frequency transfer by the passive all-optical synchronization. For this purpose, we actively lock the repetition rate of the master laser by using a PZT for cavity-length stabilization. The standard-deviation fluctuations of repetition rate are 0.5 mHz and 0.6 Hz with and without locking, respectively, as shown in Fig. \ref{fig6}(b). With the help of synchronization, the initial 2.0-Hz fluctuation of repetition rate for the slave laser is substantially reduced to 0.5 mHz as well, which indicates a high-precision frequency transfer between the two synchronized fiber lasers. 

\begin{figure}[b!]
\centering
\includegraphics[width=0.7\columnwidth]{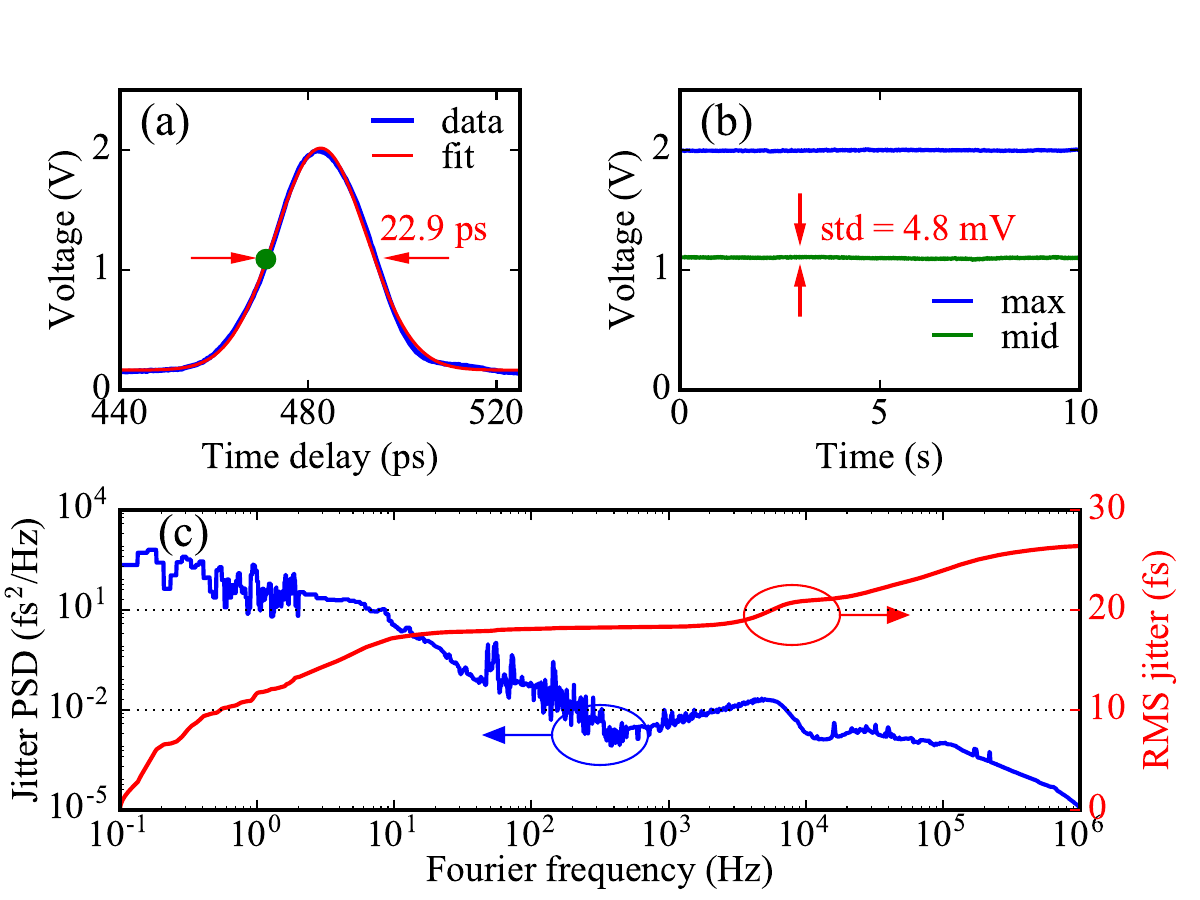}
\caption{(a) Cross-correlation trace between the two synchronized pulses. The green dot at the half amplitude indicates the position to measure the timing jitter. (b) The voltage fluctuations within a 1-MHz bandwidth at the maximum and middle points at the correlation trace. (c) Timing jitter power spectral density (left) and the integrated timing jitter in Fourier domain (right).}
\label{fig7}
\end{figure}

Finally, to quantitatively evaluate the short-term stability of passive synchronization, the timing jitter between the two-color pulses is measured by using the optical cross-correlation technique. As shown in Fig. \ref{fig1}(c), the two synchronized pulses are spatially combined by a WDM. One of the two beams can be delayed by a linear translation stage (Thorlabs LTS300/M) to control the temporal overlap. The mixed beam is then focused into a 25-mm periodically-poled lithium niobate (PPLN) crystal for sum-frequency generation (SFG). The SFG signal at 634 nm is steered through a spectral filer and detected by a gain-variable photodiode (Thorlabs PDA100A2). The output electronic signal passes a 1-MHz low-pass filter for subsequent analysis. Figure \ref{fig7}(a) shows the cross-correlation trace measured by remotely scanning the translational stage. The cross-correlation bandwidth $\tau_3$ is measured to be 22.9 ps, which is in agreement with the value calculated by the formula $\tau_3 = \sqrt{\tau_1^2 + \tau_2^2}$, where $\tau_{1,2}$ denote the durations of the two synchronized pulses. Therefore, the timing jitter should be negligible in our synchronization system. 

More rigorous deduction of timing jitter can be obtained by the Fourier analysis of the cross-correlation signal at its half maximum. In this position, the temporal variation linearly translates into the intensity fluctuation. The related slope is about 112 mV/ps in the experiment. We record the temporal evolution of the SFG signal at the half-intensity level as shown in Fig. \ref{fig7}(b). Consequently, the 4.8-mV variation without correction of detector noise and intrinsic intensity fluctuation indicates that the resultant timing jitter has an upper bound of 42 $\pm$ 9 fs. The uncertainty is due to the limited vertical resolution of the oscilloscope. Figure \ref{fig7}(c) shows the power spectral density from 0.1 Hz to 1 MHz in the Fourier domain, which is measured by a signal source analyzer (AnaPico APPH6040). The integrated root-mean-square timing jitter of 26 fs is obtained, which reveals tight synchronization enabled by our novel master-slave scheme.

\section{Conclusions}
In summary, we have presented an all-polarization-maintaining synchronized fiber laser system, which was comprised of two independent Yr-doped and Er-doped mode-locked fiber lasers in the master-slave configuration. The robust self-starting and long-term stable operation was made possible by leveraging the nonlinear amplifying loop mirror as the high-speed intensity modulator and fast artificial saturable absorber for synchronous mode-locking in the full polarization-maintaining structure. The tolerance of the cavity-length mismatch was about 800 $\mu$m, which could be further improved by increasing the injected pulse energy. The large capture range enabled a passive locking of the relative repetition rate over 12 hours, notably with a frequency error of $5 \times 10^{-10}$. The high-precision frequency transfer between the synchronized lasers was also verified by combing active locking of the master laser and passive synchronization of the slave laser. The measured timing jitter was 26 fs in a Fourier frequency range from 0.1 Hz to 1 MHz. The timing jitter can be reduced with the help of shortening the synchronized pulse durations by cavity dispersion management, or/and resorting to low-bandwidth active jitter cancellation to remove the slow timing drifts \cite{Yoshitomi2005, Hudson2006}. Benefiting from the full polarization-maintaining architecture, the presented passive all-optical synchronization system would be attractive in applications beyond laboratorial operations, such as remote clock stabilization and distribution \cite{Foreman2007}. Additionally, recent studies have demonstrated the possibility for the passive synchronization technique to transfer the offset frequency of a mode-locked laser besides the repetition frequency \cite{Betz2004, Kuse2012}. This observation may not only arouse interests for fundamental investigation of the involved nonlinear coupling mechanism, but also promote applications in duplicating stability and accuracy among frequency combs with alleviative requirement of sophisticated active feedback.

\section*{Funding.} Program for Professor of Special Appointment (Eastern Scholar) at Shanghai Institutions of Higher Learning, National Natural Science Foundation of China (11727812), Science and Technology Innovation Program of Basic Science Foundation of Shanghai (18JC1412000).


\end{document}